\documentclass[conference]{IEEEtran}
\IEEEoverridecommandlockouts
\usepackage{cite}
\usepackage{url}
\usepackage{amsmath,amssymb,amsfonts}
\usepackage{algorithmic}
\usepackage{graphicx}
\usepackage{textcomp}
\usepackage{xcolor}
\usepackage{latexsym, physics}

\usepackage{tikz}
\usetikzlibrary{matrix}

\usepackage{tikzit}

\tikzstyle{none-small}=[fill=none, draw=none, shape=circle, tikzit category=misc, tikzit shape=circle, tikzit fill=none, font={\footnotesize}]
\tikzstyle{none-small-gray}=[fill=none, draw=none, shape=circle, text=gray, tikzit category=misc, tikzit shape=circle, tikzit fill=none, font={\footnotesize}]
\tikzstyle{gate}=[shape=rectangle, text height=1ex, text depth=0.25ex, yshift=0.5mm, fill=white, draw=black, minimum height=3mm, yshift=-0.5mm, minimum width=3mm, font={\footnotesize}, tikzit category=circuit]
\tikzstyle{meter}=[shape=rectangle, text height=1ex, text depth=0.25ex, yshift=0.5mm, fill=white, draw=black, minimum height=3mm, yshift=-0.5mm, minimum width=3mm, font={\footnotesize}, tikzit category=circuit, text width=4.5mm, label={{[shift={(0,-1.15)}]\metersymb}}]
\tikzstyle{big gate}=[shape=rectangle, text height=1.5ex, text depth=0.25ex, yshift=0.5mm, fill=white, draw=black, minimum height=10mm, yshift=-0.5mm, minimum width=5mm, font={\normalsize}, tikzit category=circuit]
\tikzstyle{long gate}=[shape=rectangle, text height=1ex, text depth=0.25ex, yshift=0.5mm, fill=white, draw=black, minimum height=3mm, yshift=-0.5mm, minimum width=5mm, font={\footnotesize}, tikzit category=circuit]
\tikzstyle{Z dot}=[inner sep=0mm, minimum size=2mm, shape=circle, draw=black, fill={rgb,255: red,221; green,255; blue,221}, tikzit category=zx]
\tikzstyle{Z phase dot}=[minimum size=5mm, font={\footnotesize\boldmath}, shape=rectangle, rounded corners=2mm, inner sep=1mm, outer sep=-2mm, scale=0.8, tikzit shape=circle, draw=black, fill={rgb,255: red,221; green,255; blue,221}, tikzit draw=blue, tikzit category=zx]
\tikzstyle{X dot}=[Z dot, shape=circle, draw=black, fill={rgb,255: red,255; green,136; blue,136}, tikzit category=zx]
\tikzstyle{X phase dot}=[Z phase dot, tikzit shape=circle, tikzit draw=blue, fill={rgb,255: red,255; green,136; blue,136}, font={\footnotesize\boldmath}, tikzit category=zx]
\tikzstyle{hadamard}=[fill=yellow, draw=black, shape=rectangle, inner sep=0.6mm, minimum height=1.5mm, minimum width=1.5mm, tikzit category=zx]
\tikzstyle{paulibox}=[fill={rgb,255: red,221; green,221; blue,255}, draw=black, shape=rectangle, inner sep=0.6mm, minimum height=5mm, minimum width=5mm, font={\footnotesize}, text height=1.5ex, text depth=0.25ex, tikzit category=zx]
\tikzstyle{vertex}=[inner sep=0mm, minimum size=1mm, shape=circle, draw=black, fill=black, tikzit category=misc]
\tikzstyle{vertex set}=[inner sep=0mm, minimum size=1mm, shape=circle, draw=black, fill=white, font={\footnotesize\boldmath}, tikzit category=misc]
\tikzstyle{small black dot}=[fill=black, draw=black, shape=circle, inner sep=0pt, minimum width=1.2mm, tikzit category=circuit]
\tikzstyle{cnot ctrl}=[fill=black, draw=black, shape=circle, inner sep=0pt, minimum width=1.2mm, tikzit category=circuit]
\tikzstyle{cnot targ}=[fill=white, draw=white, shape=circle, tikzit category=circuit, label={center:$\oplus$}, inner sep=0pt, minimum width=2.1mm, tikzit fill={rgb,255: red,102; green,204; blue,255}, tikzit draw=black]
\tikzstyle{ket}=[fill=white, draw=black, shape=regular polygon, regular polygon sides=3, regular polygon rotate=-30, scale=0.7, inner sep=1pt, tikzit category=circuit, tikzit shape=rectangle, tikzit fill=green]
\tikzstyle{bra}=[fill=white, draw=black, shape=regular polygon, regular polygon sides=3, regular polygon rotate=30, scale=0.7, inner sep=1pt, tikzit category=circuit, tikzit shape=rectangle, tikzit fill=red]
\tikzstyle{scalar}=[shape=rectangle, text height=1.5ex, text depth=0.25ex, yshift=0.5mm, fill=white, draw=black, minimum height=5mm, yshift=-0.5mm, minimum width=5mm, font={\normalsize}]
\tikzstyle{clabel}=[fill=white, draw=none, shape=rectangle, tikzit fill={rgb,255: red,56; green,255; blue,242}, font={\footnotesize}, inner sep=1pt, tikzit category=labels]
\tikzstyle{empty diagram}=[draw={gray!40!white}, dashed, shape=rectangle, minimum width=1cm, minimum height=1cm, tikzit category=misc]
\tikzstyle{cluster small}=[fill=none, thick, draw={rgb,255: red,0; green,128; blue,128}, shape=circle, tikzit category=misc, tikzit shape=circle, minimum size=1.5mm,  inner sep=0.3mm, tikzit fill=white, tikzit draw={rgb,255: red,0; green,128; blue,128}, font={\footnotesize}]
\tikzstyle{cluster}=[fill=none, thick, draw={rgb,255: red,0; green,128; blue,128}, shape=circle, tikzit category=misc, tikzit shape=circle, minimum size=3.5mm, inner sep=0pt, tikzit fill=white, tikzit draw={rgb,255: red,0; green,128; blue,128}, font={\footnotesize}]
\tikzstyle{cluster big}=[fill=none, thick, draw={rgb,255: red,0; green,128; blue,128}, shape=circle, tikzit category=misc, tikzit shape=circle, minimum size=4.5mm, text width=2mm, inner sep=0pt, tikzit fill=white, tikzit draw={rgb,255: red,0; green,128; blue,128}, font={\footnotesize}]

\tikzstyle{hadamard edge}=[-, dashed, dash pattern=on 2pt off 0.5pt, thick, draw={rgb,255: red,68; green,136; blue,255}]
\tikzstyle{box edge}=[-, dashed, dash pattern=on 2pt off 0.5pt, thick, draw={rgb,255: red,203; green,192; blue,225}]
\tikzstyle{brace edge}=[-, tikzit draw=blue, decorate, decoration={brace,amplitude=1mm,raise=-1mm}]
\tikzstyle{diredge}=[->]
\tikzstyle{double edge}=[-, double, shorten <=-1mm, shorten >=-1mm, double distance=2pt]
\tikzstyle{gray edge}=[-, {gray!70!white}, thick]
\tikzstyle{pointer edge}=[->, very thick, gray]
\tikzstyle{boldedge}=[-, line width=1.2pt, shorten <=-0.17mm, shorten >=-0.17mm]
\tikzstyle{boldedge red}=[-, line width=1.4pt, shorten <=-0.17mm, shorten >=-0.17mm, draw=red, tikzit draw=red]

\newcommand\bovermat[2]{%
	\makebox[0pt][l]{$\smash{\overbrace{\phantom{%
					\begin{matrix}#2\end{matrix}}}^{#1}}$}#2}

\begin{document}
\title{Teaching quantum computing to computer science students: Review of a hands-on quantum circuit simulation practical}

\author{\IEEEauthorblockN{1\textsuperscript{st} Florian Krötz}
\IEEEauthorblockA{\textit{MNM-Team} \\
\textit{Ludwig-Maximilians-}\\
\textit{Universität München}\\
Munich, Germany \\
\textit{florian.kroetz@nm.ifi.lmu.de}}
\and
\IEEEauthorblockN{2\textsuperscript{nd} Xiao-Ting Michelle To}
\IEEEauthorblockA{\textit{MNM-Team} \\
\textit{Ludwig-Maximilians-}\\
\textit{Universität München}\\
Munich, Germany \\
\textit{michelle.to@nm.ifi.lmu.de}}
\and
\IEEEauthorblockN{3\textsuperscript{rd} Korbinian Staudacher}
\IEEEauthorblockA{\textit{MNM-Team} \\
\textit{Ludwig-Maximilians-}\\
\textit{Universität München}\\
Munich, Germany \\
\textit{staudacher@nm.ifi.lmu.de}}
\and
\IEEEauthorblockN{4\textsuperscript{th} Dieter Kranzlmüller}
\IEEEauthorblockA{\textit{MNM-Team} \\
\textit{Ludwig-Maximilians-}\\
\textit{Universität München}\\
Munich, Germany \\
\textit{kranzlmueller@ifi.lmu.de }}
}
\maketitle

\begin{abstract}
We present a practical course targeting graduate students with prior knowledge of the basics of quantum computing. The practical aims to deepen students' understanding of fundamental concepts in quantum computing by implementing quantum circuit simulators. Through hands-on experience, students learn about different methods to simulate quantum computing, including state vectors, density matrices, the stabilizer formalism, and matrix product states.

By implementing the simulation methods themselves, students develop a more in-depth understanding of fundamental concepts in quantum computing, including superposition, entanglement, and the effects of noise on quantum systems.

This hands-on experience prepares students to do research in the field of quantum computing and equips them with the knowledge and skills necessary to tackle complex research projects in the field.
In this work, we describe our teaching approach and the structure of our practical, and we discuss evaluations and lessons learned.

\end{abstract}


\section{Introduction}
\label{introduction}
The growing importance of quantum computing has created a need for skilled computer scientists to work in this field. However, teaching quantum computing to computer science students is a challenging task, as they often lack a fundamental understanding of quantum mechanics, including concepts such as superposition and entanglement.

The target audience for the practical consists of students who have successfully completed an introductory lecture on quantum computing~\cite{homeister2013quantum, Nielsen_Chuang_2010, aaronson2022qclec, LMU2025}. The lecture provides a solid foundation in the fundamentals of quantum computing, but students often require additional support to develop a solid understanding of the subject.

To address this knowledge gap, we designed a Quantum Computing Practical, a hands-on learning experience aimed at deepening students' understanding of quantum computing principles.
The practical should not replace an introductory lecture; it builds on the lecture. The primary goals are to equip students with a more in-depth understanding of quantum computing to write and do research in the field.

Our practical course is delivered in an intensive one-week block format. This structure encourages a deeper immersion into the quantum simulation topic and allows students to complete more ambitious projects within a condensed timeframe.

The structure of this paper is as follows:
Section~\ref{RelatedWork} gives an overview of existing teaching approaches for computer science students and motivates why we decided for the teaching approach and learning goals described in Section~\ref{TeachingApproach}, while Section~\ref{Course} outlines how we implemented the described approach. In Section~\ref{Evaluation}, we present the evaluation of the practical, provide conclusions and an outlook on future work.

\section{Related Work}
\label{RelatedWork}
A growing number of educational initiatives have been developed to support teaching quantum computing~\cite{10821135, Carrascal2021-fl, 9951184}, often leveraging interactive tools and simulators~\cite{zuse_tn_workshop, eu_tn_workshop_phd}.
Our approach is more aligned with constructivist learning paradigms, where students gain more in-depth understanding by constructing the quantum circuit simulators themselves.

There is a growing number of simulators for quantum computing~\cite{Gangapuram2024-ot}. 
One goal of this practical is that students can make a well-informed decision which simulator fits best for their use case. 

Several teaching platforms or toolkits have emerged to support quantum computing education. 
For instance, IBM Quantum Learning~\cite{ibm_quantum_learning}, Qiskit~\cite{qiskit2024}, and Cirq~\cite{CirqDevelopers_2025} provide cloud-based access to quantum hardware and simulators. 
However, while these platforms are good for demonstrating quantum algorithms, they often abstract away implementation details.

In contrast, our course emphasizes building simulators from scratch, helping students internalize core computational models such as statevectors~\cite{Nielsen_Chuang_2010}, density matrices~\cite{Nielsen_Chuang_2010}, the stabilizer formalism~\cite{Aaronson_2004, aaronson2022qclec, Bravyi2019simulationofquantum}, and matrix product states~\cite{PhysRevLett.91.147902, PhysRevX.10.041038}.

Some courses and workshops have incorporated tensor network methods into their curriculum, such as the tutorials provided at the Quantum Tensor Network Summer School~\cite{eu_tn_workshop_phd} targeting PhD students, or targeting students with more prior knowledge~\cite{zuse_tn_workshop}.

Our practical gives a comprehensive introduction to tensor networks for computer science students.
Linear algebra lectures for computer science students usually do not introduce tensors~\cite{la_hamburg, la_tum, la_mit}. 
Some lectures about deep learning do introduce tensors~\cite{deeplearing_lecture}, but since these are advanced lectures, not every computer science student attended such a lecture. 
However, not having the prior knowledge about tensors and tensor networks makes reading literature on the matrix product state~\cite{PhysRevLett.91.147902, PhysRevX.10.041038} harder to understand.

\section{Teaching approach}
\label{TeachingApproach}
Computer scientists in quantum computing often use simulators to understand and analyze quantum systems. With multiple simulation methods available, it is essential to know the strengths and weaknesses of each method to choose the right simulator and operate it correctly.

\subsection{Pedagogical Concept}
The practical course goes beyond teaching simulators, focusing on developing an in-depth understanding of quantum computing fundamentals through hands-on experience by implementing and optimizing various simulation techniques and exploring noise models. 

Throughout the course, supervisors have been present to provide guidance and support, ensuring that students have access to expert knowledge and advice whenever needed. Students have also been encouraged to collaborate and help each other.

For the practical, we use the gate model as the foundation for representing quantum circuits. 
Since the focus of the practical is understanding the fundamentals in quantum computing, we decided not to use an established implementation of the gate model like Qiskit~\cite{qiskit2024}, Cirq~\cite{CirqDevelopers_2025}, or OpenQASM~\cite{openqasm3}, because there is a lot of existing software that abstracts away the underlying quantum computing concepts. 
For example, Qiskit implements many high-level functions where the user does not have to work at the gate level.
We want the students to think about quantum computing on the gate level.
OpenQASM has many complex features, like registers or loops, that make it harder to parse.
Therefore, we developed a simple representation of quantum circuits, inspired by the representation used in~\cite{Aaronson_2004}. 
The core idea here is that a quantum circuit can be represented as a list of gates operating on a set of qubits.
By using a simple representation, students have to focus on the underlying principles.
 
\subsection{Learning Goals}\label{subsec:learning-goals}
By quantum circuit simulation, we mean to determine the output state $\ket{\psi'}$ obtained by applying a unitary operator $U$, which represents one or more quantum gates, to an initial state $\ket{\psi}$:
\[
\ket{\psi'} =  U \ket{\psi}.
\]

For all methods, a central learning goal is the understanding how a quantum state $\ket{\psi}$ is represented,  what are strength and weaknesses of this representation and how it is modified when applying a unitary $U$.
The primary learning goals for the practical are:
\begin{itemize}
    \item A more in-depth understanding of superposition and entanglement is built, since students learn about 4 different methods of how to represent a quantum state on a classical computer.
    \item A solid understanding of the various simulation techniques, enabling students to make informed decisions about which simulator to use.
    \item An intuition for the non-local nature of quantum computing, which means that quantum circuits are generally not separable and representing a general quantum state is exponential hard.
\end{itemize}
The main learning goals for the four chosen methods are:
\begin{itemize} 
    \item State Vector: When representing a quantum state as a State vector, the memory needed grows exponential $O(2^n)$, where n is the number of qubits.
    Single-qubit operations may change every entry of the state vector, illustrating non-locality. Circuits are represented as a list of gates, but every entry in the state vector needs to be touched, demonstrating non-separability because of the non-local nature of quantum mechanics.
    \item Density Matrix with Noise:  When representing a quantum state as a Density Matrix, the memory needed grows exponential $O(2^{2n})$ and needs quadratic more memory than a state vector. An advantage is that noise can be modeled. Another learning goal is what error types are there and how they are modeled.
    \item Stabilizer Formalism: The main limitation is that only stabilizer states can be represented, but they can be efficient represented with $O(n^2)$.
    Quantum systems can be simulated efficiently, but their quantum nature alone does not ensure their usefulness.
    \item Matrix Product State (MPS): In an MPS, the space complexity depends on the degree of entanglement in the system. Students learn how to decompose an entangled system, building a better understanding of entanglement and showing what makes quantum systems hard to simulate. 
    
    A small but worth mentioning learning goal here is that students get an intuition for what tensors are. 
\end{itemize}

\section{Course}
\label{Course}
\subsection{Course Description}
The official description is:
"The topic of the practical course is simulation of quantum circuits. Participating students are learning different methods for classical simulation of quantum circuits. Based on those, they implement and optimize their own simulator in groups of two."

The course is structured in two stages: There is a worksheet phase where the students work on worksheets. For each simulation method, there is one worksheet.
The second phase is the project phase, where the students implement their own simulator.
Our practical course is delivered in an intensive seven-day block format. This structure encourages a deeper immersion into the quantum circuit simulation topic and allows students to complete more ambitious projects within a condensed time frame.

\subsection{Circuit representation}
The quantum circuit representation used in our practical is called QCP (Quantum Computing Practical).
The representation dispenses with advanced features such as multiple quantum registers, classical registers, and conditional gates.

QCP circuits are represented in text files that need to be in the following form.
The first line contains the number of qubits.
Each of the following lines represents exactly one gate in the format: \texttt{gateName [parameter] [controlqubit(s)] targetqubit}.
The parameters \texttt{controlqubit} and \texttt{targetqubit} are natural numbers between $0$ and the number of qubits$-1$; \texttt{parameter} is a decimal number, optionally multiplied by a fraction of $\pi$.

For example, a two-qubit circuit that produces a Bell state by first applying a Hadamard gate on the first qubit and then a CX gate, with the control qubit being the first qubit and the target qubit the second, is described as follows:\\
\texttt{2\\ h 0\\ cx 0 1}

We provide the students with a reference implementation in Python and C. 
The representation is deliberately kept simple such that writing their own parser in another language is not difficult.

\subsection{Worksheets}
The worksheet phase of the practical contains six worksheets that the participants are working on. 
The first worksheet is a recap, and worksheets 2–5 introduce the simulation methods. 
The time scope is approximately half a day for each  worksheet; in sum three days for the worksheet phase.

\subsubsection{Recap}
The first worksheet is a recap where lecture topics important for the practical course are covered. Thereby, the worksheets have a special focus on unitary operations, tensor products, how to define quantum gates in Dirac notation and as matrices, and density matrices.

\subsubsection{Statevector}
This worksheet introduces statevector simulation and familiarizes students with the QCP circuit representation. In the sheet, students implement a not optimized state vector simulator in python.

The first task is to initialize a state vector $\ket{\psi}$ with $n$ qubits in the state $\ket{0}^{\otimes n}$. 
Here, the students learn that the state vector has to grow exponential with $O(2^n)$.
The next task is to implement gates that modify the state vector $\ket{\psi}$.
We ask to create a unitary matrix $U = I \otimes \cdots \otimes U_i \otimes \cdots \otimes I$ by multiplying, using the Kronecker product, identity matrices $I$ and $U_i$ being a single-qubit unitary acting on qubit $i$ or a CNOT unitary unitary that is constructed by
\[
	 CNOT = \underbrace{\ket{0}\bra{0}}_{\text{control}} \otimes \underbrace{I}_{\text{target}} + \underbrace{\ket{1}\bra{1}}_{\text{target}} \otimes \underbrace{U}_{\text{target}}.
\]
If the control and target qubit are not next to each other Kronecker products of identity matrices have to be placed between the control and target qubit.
This matrix is then multiplied on the state $\ket{\psi'} =  U \ket{\psi}$.

We ask the students to validate their implementation by using circuits that produce a Bell state or a GHZ state, and we ask them to visualize the result.
Through validating their implementation and visualizing the results, the students get a good understanding of how a state vector represents a quantum state.

Finally, the students implement measurements on single qubits using the measurement operators $M_0=\ket{0}\bra{0}$ and $M_1=\ket{1}\bra{1}$.
The measurement operators are applied to the state vector like single-qubit gates. After the measurement operator has been applied to the state vector, the state vector needs to be normalized with
\[
 \dfrac{M_m\ket{\psi}}{\sqrt{\bra{\psi}M_m^\dagger M_m\ket{\psi}}}.
\]

\subsubsection{Density Matrix}
The structure of this sheet is similar to the previous sheet on statevector simulation. In the sheet, students implement a not-optimized density matrix simulator in python.

First, a quantum state is initialized as a density matrix $\rho = \ket{\psi}\bra{\psi}$. For simplicity, we only consider pure states here.
Then gates and measurements with $\rho'= U \rho U^\dagger$ are implemented. 
Here, the same $U$ as in the state vector sheet can be used. 
Then the following errors are introduced and how they are modeled: bit flip, phase flip, bit phase flip, depolarization, and amplitude damping~\cite{Nielsen_Chuang_2010}.

\subsubsection{Stabilizer Formalism}
This worksheet introduces simulation using the stabilizer formalism. This is a pen-and-paper-only sheet. 
This sheet closely follows Lecture 28 of~\cite{aaronson2022qclec}. We start with the definition of stabilizer states.
To represent a quantum state using the stabilizer formalism, we first need to introduce stabilizer states. A state $\ket{\psi}$ is called a stabilizer state when $P \ket{\psi} = \ket{\psi}$ for Pauli matrices $P \in \{Z,X,Y,I\}$. 
Systems with multiple qubits are stabilized by Pauli strings. A Pauli string is a product of Pauli matrices, denoted as $P_1 \otimes \cdots \otimes P_n$, where each $P_i \in P$, $i\in\{1,\dots,n\}$, is a Pauli matrix.
Pauli strings can be efficiently represented in a so-called tableau: 
\begin{align*}
    \left(
    \begin{array}{ccc|ccc}
        x_{1,1}     & \hdots & x_{1,n}     & z_{1,1}     & \hdots  & z_{1,n} \\
        \vdots      & \ddots & \vdots      & \vdots      & \ddots  & \vdots  \\
        x_{n,1}     & \hdots & x_{n,n}     & z_{n,1}     & \hdots  & z_{n,n} \\ 
    \end{array}
    \right),
    \begin{tabular}{c|cc} 
    P & $x_i$ & $z_i$ \\\hline
    I & 0 & 0 \\ 
    X & 1 & 0 \\ 
    Z & 0 & 1 \\
    Y & 1 & 1
\end{tabular}
\end{align*}
The binary matrix on the left is the tableau, and the right table describes how Pauli matrices are encoded in the tableau. The entries in the tableau are binary. 
A tableau has $n$ rows, with each of them representing a Pauli string.
The Pauli strings represented by the tableau stabilize the quantum state.
This representation of a state is much denser than a state vector
This efficient representation of a state comes with the cost that only stabilizer states can be represented. 

When simulating using the stabilizer formalism, only Clifford gates $H,S,CNOT$ can be applied. The Clifford gates can be realized as simple bit operations on a stabilizer tableau~\cite{aaronson2022qclec}.
A nice side effect here is that the participants gain a better understanding of the Hadamard gate: 
"The Hadamard gate swaps the X and Z basis, and in the stabilizer formalism, one just needs to swap the stabilizers for X and Z."~\cite{aaronson2022qclec}

\subsubsection{Matrix Product State (MPS)}
The last worksheet before the project sheet introduces matrix product state (MPS) simulation. This is a pen-and-paper-only sheet. 
Since most computer science curricula do not introduce tensors and the tensor diagram notation, we introduce it with an example most computer science students know: the matrix product. 
In this example, the matrices $A$, $B$ and $C$ are rank-2 tensors. 
The multiplication of
\begin{align*}
    C_{i,j} = \sum_k A_{i,k} B_{k,j} \ \Longleftrightarrow \ \input{mmcontraction.tikz}
\end{align*}
is an example of tensor contraction.
We then introduce the MPS following~\cite{PhysRevX.10.041038}.  

Every qubit is represented by a tensor network of the following form:
\[
    \input{tensormps.tikz}
\]
The qubits are ordered in a line, where only adjacent qubits can be entangled.
Since most computer science students are struggling to imagine a rank-3 tensors, we give them a visual intuition of how such a tensor in an MPS looks like. 
These tensors are two matrices:
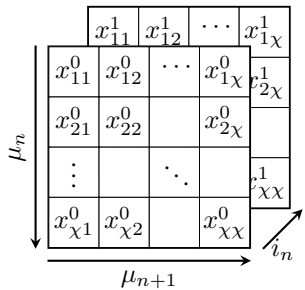
\[
    \begin{tikzpicture}[auto matrix/.style={matrix of nodes,
		draw,thick,inner sep=0pt,
		nodes in empty cells,column sep=-0.2pt,row sep=-0.2pt,
		cells={nodes={minimum width=1.9em,minimum height=1.9em,
				draw,very thin,anchor=center,fill=white,
				execute at begin node={%
					$\vphantom{x_|}\ifnum\the\pgfmatrixcurrentrow<3
					\ifnum\the\pgfmatrixcurrentcolumn<3
					x^{#1}_{\the\pgfmatrixcurrentrow\the\pgfmatrixcurrentcolumn}
					\else 
					\ifnum\the\pgfmatrixcurrentcolumn=4
					x^{#1}_{\the\pgfmatrixcurrentrow\chi}
					\fi
					\fi
					\else
					\ifnum\the\pgfmatrixcurrentrow=4
					\ifnum\the\pgfmatrixcurrentcolumn<3
					x^{#1}_{\chi\the\pgfmatrixcurrentcolumn}
					\else
					\ifnum\the\pgfmatrixcurrentcolumn=4
					x^{#1}_{\chi\chi}
					\fi 
					\fi
					\fi
					\fi
                    \ifnum\the\pgfmatrixcurrentrow\the\pgfmatrixcurrentcolumn=13
					\cdots
					\fi
                    \ifnum\the\pgfmatrixcurrentrow\the\pgfmatrixcurrentcolumn=31
					\vdots
					\fi
					\ifnum\the\pgfmatrixcurrentrow\the\pgfmatrixcurrentcolumn=33
					\ddots
					\fi$
				}
	}}}]
	\matrix[auto matrix=1,xshift=1.5em,yshift=1.5em](maty){
		& & & \\
		& & & \\
		& & & \\
		& & & \\
	};
	 \matrix[auto matrix=0](matx){
		& & & \\
		& & & \\
		& & & \\
		& & & \\
	};
	\draw[thick,-stealth] ([xshift=1ex]matx.south east) -- ([xshift=1ex]maty.south east)
	node[midway,below] {$i_n$};
	\draw[thick,-stealth] ([yshift=-1ex]matx.south west) -- 
	([yshift=-1ex]matx.south east) node[midway,below] {$\mu_{n+1}$};
	\draw[thick,-stealth] ([xshift=-1ex]matx.north west)
	-- ([xshift=-1ex]matx.south west) node[midway,above,rotate=90] {$\mu_n$};
\end{tikzpicture}
\]
The front matrix corresponds to the case the qubit is $\ket{0}$, and the back matrix corresponds to $\ket{1}$.

The operation of a single qubit gate is given by 
\[
   M'(n)^{i_n'}_{\mu_{n-1}\mu_n} = \sum_{i_n}U_{i_n'i_n}M(n)^{i_n}_{\mu_{n-1}\mu_n}.
\]
In the tensor network diagram, this means plugin the single qubit gate unitary on the open index of the tensor of the corresponding qubit.

Applying the two-qubit gate in an MPS is more complex. We introduce it using the tensor network diagram:
\[
    \input{2qbit2.tikz}
\]
A two-qubit unitary is plugged on the open indices of two adjacent qubits. 
This two-qubit unitary is represented as a rank-4 tensor. 
For such a rank-4 tensor, we give the participants the CNOT
\[
CX^{i,j}_{k,l}:
\begin{array}{l cccc}\\
    CX^{00} = & (\bovermat{k=0}{1 & 0} &\bovermat{k=1}{0 & 0}) \\
    CX^{01} = & (0 & 1 & 0 & 0) \\
    CX^{10} = & (0 & 0 & 0 & 1) \\
    CX^{11} = & (0 & 0 & 1 & 0) \\ 
    {}_{l=}& {}_{0} & {}_{1} & {}_{0} & {}_{1}
\end{array}
\]
as an example to build up an understanding of such an operation.
After the contraction, the new rank-4 tensor $T'$ needs to be separated into two rank-3 tensors.
To obtain two 3-dimensional tensors again, the 4-dimensional tensor is first interpreted as a 2-dimensional matrix (analogous to the 2D representation of the $CX$ gate on the first page of the exercise sheet: Index $i_n'$ and $\mu_{n-1}$ form one dimension, index $i_{n+1}'$ and $\mu_{n+1}'$ the other).
A \textit{singular value decomposition (SVD)} can now be applied to this 2-dimensional matrix. 
It decomposes a matrix into three components $XSY$. 
$X$ and $Y$ are both unitary matrices, and $S$ is a diagonal matrix. 
The diagonal entries are called \textit{singular values}.
By multiplying the singular values to $X$ and interpreting the 2-dimensional matrices as 3-dimensional tensors again (i.e., splitting the $X$ matrix horizontal in the middle into a $0$ and a $1$ part and the $Y$ matrix vertical), the modified MPS is obtained.
The implementation of this procedure with NumPy~\cite{numpy} is trivial. 
Going through this process with pen and paper illustrates how the simulation process in an MPS works. 
In our case, the process starts with the state $\ket{0+}$
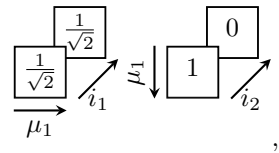
\[
    \begin{tikzpicture}[auto matrix/.style={matrix of nodes,
        draw,thick,inner sep=0pt,
        nodes in empty cells,column sep=-0.2pt,row sep=-0.2pt,
        cells={nodes={minimum width=1.9em,minimum height=1.9em,
                draw,very thin,anchor=center,fill=white,
                execute at begin node={%
                    $\vphantom{x_|}\ifnum\the\pgfmatrixcurrentrow<4
                    \ifnum\the\pgfmatrixcurrentcolumn<4
                    {\frac{1}{\sqrt{2}}}
                    \else 
                    \ifnum\the\pgfmatrixcurrentcolumn=5
                    {\frac{1}{\sqrt{2}}}
                    \fi
                    \fi
                    \else
                    \ifnum\the\pgfmatrixcurrentrow=5
                    \ifnum\the\pgfmatrixcurrentcolumn<4
                    {\frac{1}{\sqrt{2}}}
                    \else
                    \ifnum\the\pgfmatrixcurrentcolumn=5
                    {\frac{1}{\sqrt{2}}}
                    \fi 
                    \fi
                    \fi
                    \fi  
                    \ifnum\the\pgfmatrixcurrentrow\the\pgfmatrixcurrentcolumn=14
                    \cdots
                    \fi
                    \ifnum\the\pgfmatrixcurrentrow\the\pgfmatrixcurrentcolumn=41
                    \vdots
                    \fi
                    \ifnum\the\pgfmatrixcurrentrow\the\pgfmatrixcurrentcolumn=44
                    \ddots
                    \fi$
                }
    }}}]
    \matrix[auto matrix=1,xshift=1.5em,yshift=1.5em](maty){
        \\
    };
    \matrix[auto matrix=0](matx){
        \\
    };
    \draw[thick,-stealth] ([xshift=1ex]matx.south east) -- ([xshift=1ex]maty.south east)
    node[midway,below] {$i_1$};
    \draw[thick,-stealth] ([yshift=-1ex]matx.south west) -- 
    ([yshift=-1ex]matx.south east) node[midway,below] {$\mu_1$};
    
\end{tikzpicture}
\begin{tikzpicture}[auto matrix/.style={matrix of nodes,
        draw,thick,inner sep=0pt,
        nodes in empty cells,column sep=-0.2pt,row sep=-0.2pt,
        cells={nodes={minimum width=1.9em,minimum height=1.9em,
                draw,very thin,anchor=center,fill=white,
                execute at begin node={%
                    $\vphantom{x_|}\ifnum\the\pgfmatrixcurrentrow<4
                    \ifnum\the\pgfmatrixcurrentcolumn<4
                    {#1}
                    \else 
                    \ifnum\the\pgfmatrixcurrentcolumn=5
                    {#1}
                    \fi
                    \fi
                    \else
                    \ifnum\the\pgfmatrixcurrentrow=5
                    \ifnum\the\pgfmatrixcurrentcolumn<4
                    {0}
                    \else
                    \ifnum\the\pgfmatrixcurrentcolumn=5
                    {0}
                    \fi 
                    \fi
                    \fi
                    \fi  
                    \ifnum\the\pgfmatrixcurrentrow\the\pgfmatrixcurrentcolumn=14
                    \cdots
                    \fi
                    \ifnum\the\pgfmatrixcurrentrow\the\pgfmatrixcurrentcolumn=41
                    \vdots
                    \fi
                    \ifnum\the\pgfmatrixcurrentrow\the\pgfmatrixcurrentcolumn=44
                    \ddots
                    \fi$
                }
    }}}]
    \matrix[auto matrix=0,xshift=1.5em,yshift=1.5em](maty){
        \\
    };
    \matrix[auto matrix=1](matx){
        \\
    };
    \draw[thick,-stealth] ([xshift=1ex]matx.south east) -- ([xshift=1ex]maty.south east)
    node[midway,below] {$i_2$};
    \vphantom{\draw[thick,-stealth] ([yshift=-1ex]matx.south west) -- 
    ([yshift=-1ex]matx.south east) node[midway,below] {$\mu_2$};}
    \draw[thick,-stealth] ([xshift=-1ex]matx.north west)
    -- ([xshift=-1ex]matx.south west) node[midway,above,rotate=90] {$\mu_1$};
\end{tikzpicture},
\]
then applying a CNOT gate results in 
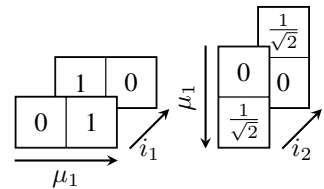
\[
    \begin{tikzpicture}[auto matrix/.style={matrix of nodes,
        draw,thick,inner sep=0pt,
        nodes in empty cells,column sep=-0.2pt,row sep=-0.2pt,
        cells={nodes={minimum width=1.9em,minimum height=1.9em,
                draw,very thin,anchor=center,fill=white}}}]
    \matrix[auto matrix=1,xshift=1.5em,yshift=1.5em](maty){
       1 & 0\\
    };
    \matrix[auto matrix=0](matx){
       0 & 1\\
    };
    \draw[thick,-stealth] ([xshift=1ex]matx.south east) -- ([xshift=1ex]maty.south east)
    node[midway,below] {$i_1$};
    \draw[thick,-stealth] ([yshift=-1ex]matx.south west) -- 
    ([yshift=-1ex]matx.south east) node[midway,below] {$\mu_1$};
    
\end{tikzpicture}
\begin{tikzpicture}[auto matrix/.style={matrix of nodes,
        draw,thick,inner sep=0pt,
        nodes in empty cells,column sep=-0.2pt,row sep=-0.2pt,
        cells={nodes={minimum width=1.9em,minimum height=1.9em,
                draw,very thin,anchor=center,fill=white}}}]
    \matrix[auto matrix=1,xshift=1.5em,yshift=1.5em](maty){
         $\frac{1}{\sqrt{2}}$ \\
          0 \\
    };
    \matrix[auto matrix=0](matx){
        0 \\
        $\frac{1}{\sqrt{2}}$ \\
    };
    \draw[thick,-stealth] ([xshift=1ex]matx.south east) -- ([xshift=1ex]maty.south east)
    node[midway,below] {$i_2$};
    \vphantom{\draw[thick,-stealth] ([yshift=-1ex]matx.south west) -- 
    ([yshift=-1ex]matx.south east) node[midway,below] {$\mu_2$};}
    \draw[thick,-stealth] ([xshift=-1ex]matx.north west)
    -- ([xshift=-1ex]matx.south west) node[midway,above,rotate=90] {$\mu_1$};
\end{tikzpicture}.
\]
Furthermore, it shows that unentangled states can be represented efficiently, and operations that introduce entanglement make the size of the tensors grow.

\subsection{Student Project}
The second stage of the practical is the student project. The students work in groups of two or three. 
For the project, they choose one of the previously introduced simulation methods and implement it in a programming language of their choice. 
The goal of the project is not just to implement a simulator, but also that the students get creative and integrate their own ideas into their simulator. 
For this, we provide them with some initial ideas:
\begin{itemize}
    \item For all simulation methods: Circuit Preprocessing, Parallelization, Cross-Entropy Benchmarking, ...
    \item Statevector Simulator: Can you apply a gate to a $n$-qubit vector without calculating a complete $n \times n$-matrix?
    \item Density Matrix Simulator: How would you implement noise in gates? Can you reproduce the behavior of real
    quantum computers using IBM error statistics, for example? You can also use error correction circuits (and
    modify them if necessary or modify your code) to correct the noise.
    \item Stabilizer Simulator: Can you omit individual T gate terms? What is the overall error induced by this?
    \item Matrix Product State Simulator: Can you choose the order of the tensor contractions so that the overall calculation runs as efficiently as possible (contraction plan)? How does the total error behave when cutting using $\chi$? Can you improve the result of the singular value decomposition? (Keyword: Canonical Matrix Product States)
\end{itemize}

\subsection{Assessment}
The assessment consists of two parts: a presentation and a discussion afterward. 
For the assessment, the students prepare a presentation. 
The students should answer the following questions in their presentation:
\begin{itemize}
    \item What simulation method did they choose and why?
    \item What programming language did they choose and why?
    \item What optimizations did they implement, and what own ideas did they integrate?
\end{itemize}
If the students can answer these questions profoundly, can explain their project decisions and give suitable answers to the questions in the discussion, they have proven that they reached the learning goals defined in Section~\ref{subsec:learning-goals}.

\section{Evaluation, Lessons Learned, and Outlook}
\label{Evaluation}
\subsection{Evaluation}
The practical has been held twice, in 2023 and 2024. In total, 28 students completed it.
Table \ref{tab:counts-year} shows the popularity of the methods. 2023 and 2024. 
At the end of each practical course, we handed out an anonymous evaluation form to the students. 
The students evaluated the practical in both years. 
The evaluation uses a scale from 1 (very good) to 5 (insufficient).
Overall, students rated the practical 1.5 (in 2023 and 2024).
The data also suggest that the practical training boosted students' interest in quantum computing, with ratings of 1.5 (2023) and 1.7 (2024) on average.
Furthermore, the evaluation results indicate that students had no problem following the content.
When asked to assess the clarity of the learning objectives, students rated it 1.8 in 2023 and 1.5 in 2024.
The results of the exam show that they do not just have fun but also reach the learning goals. 
Another goal of the practical is to prepare students to write final theses in the field of quantum computing. 
We recruited 6 participants from the practical for final theses, and all of them finished successfully. 

\begin{center}
\begin{table}[!t]
  \caption{Number of teams selecting each method by year}
  \centering
  \label{tab:counts-year}
  \begin{tabular}{l|cc}
    & \textbf{2023} & \textbf{2024} \\
    \hline
    State Vector         & 1 & 1 \\
    Density Matrix       & 1 & 4 \\
    Stabilizer Formalism & 1 & 1 \\
    Matrix Product State & 4 & 2 \\
  \end{tabular}
\end{table}
\end{center}
\subsection{Lessons Learned}
Due to historical reasons, a portion of the practical was designed as a challenge, where projects from the second phase competed against each other for the "Best Project" award. 
The intention behind this challenge was to give the students an incentive to excel in their projects and come up with new and interesting ideas.
However, the challenge turned out to be impractical, since it is difficult to design a fair competition that gives all simulation methods an equal chance of winning.
For instance, the density matrix method is known to be the slowest one but provides interesting insights into the simulation of noise in quantum computing. Therefore, students who want to win the challenge do not choose this simulation method.

\subsection{Outlook}
In the future, we aim to enhance the practical course by incorporating more hands-on experience with real hardware. The Leibniz Supercomputing Center offers access to actual quantum devices~\cite{LRZQuantum}, which we can leverage to provide students with a more immersive learning experience.

In 2024, some students worked on cross-entropy benchmarking~\cite{Boixo2018-ai} and compared their own simulator with real hardware.
It would be great to see students reproducing controversial, discussed results of the Google Sycamore Experiment~\cite{SycamoreExperiment, PhysRevLett.129.090502}, with real Quantum Computers and their own simulators.

\bibliography{QCPbibfile}

@article{Aaronson_2004,
   title={Improved simulation of stabilizer circuits},
   volume={70},
   ISSN={1094-1622},
   url={http://dx.doi.org/10.1103/PhysRevA.70.052328},
   DOI={10.1103/physreva.70.052328},
   number={5},
   journal={Physical Review A},
   publisher={American Physical Society (APS)},
   author={Aaronson, Scott and Gottesman, Daniel},
   year={2004},
   month=nov }

@article{PhysRevLett.91.147902,
  title = {Efficient Classical Simulation of Slightly Entangled Quantum Computations},
  author = {Vidal, Guifr\'e},
  journal = {Phys. Rev. Lett.},
  volume = {91},
  issue = {14},
  pages = {147902},
  numpages = {4},
  year = {2003},
  month = {Oct},
  publisher = {American Physical Society},
  doi = {10.1103/PhysRevLett.91.147902},
  url = {https://link.aps.org/doi/10.1103/PhysRevLett.91.147902}
}

@article{PhysRevX.10.041038,
  title = {What Limits the Simulation of Quantum Computers?},
  author = {Zhou, Yiqing and Stoudenmire, E. Miles and Waintal, Xavier},
  journal = {Phys. Rev. X},
  volume = {10},
  issue = {4},
  pages = {041038},
  numpages = {15},
  year = {2020},
  month = {Nov},
  publisher = {American Physical Society},
  doi = {10.1103/PhysRevX.10.041038},
  url = {https://link.aps.org/doi/10.1103/PhysRevX.10.041038}
}

@article{Bravyi2019simulationofquantum,
  doi = {10.22331/q-2019-09-02-181},
  url = {https://doi.org/10.22331/q-2019-09-02-181},
  title = {Simulation of quantum circuits by low-rank stabilizer decompositions},
  author = {Bravyi, Sergey and Browne, Dan and Calpin, Padraic and Campbell, Earl and Gosset, David and Howard, Mark},
  journal = {{Quantum}},
  issn = {2521-327X},
  publisher = {{Verein zur F{\"{o}}rderung des Open Access Publizierens in den Quantenwissenschaften}},
  volume = {3},
  pages = {181},
  month = sep,
  year = {2019}
}

@book{Nielsen_Chuang_2010, 
    place={Cambridge}, 
    title={Quantum Computation and Quantum Information: 10th Anniversary Edition}, 
    publisher={Cambridge University Press}, 
    author={Nielsen, Michael A. and Chuang, Isaac L.}, 
    year={2010}
}

@misc{aaronson2022qclec,
  author = {Scott Aaronson},
  title = {Introduction to Quantum Information Science},
  year = {2022},
  url = {https://www.scottaaronson.com/qclec.pdf},
  note = {Accessed: 2025-03-10}
}

@book{homeister2013quantum,
  title={Quantum Computing verstehen: Grundlagen - Anwendungen - Perspektiven},
  author={Homeister, M.},
  isbn={9783834818683},
  series={Computational Intelligence},
  url={https://books.google.de/books?id=s1SQkQEACAAJ},
  year={2013},
  publisher={Springer Fachmedien Wiesbaden}
}

@INPROCEEDINGS{10821135,

  author={Mjeda, Anila and Murray, Hazel},

  booktitle={2024 IEEE International Conference on Quantum Computing and Engineering (QCE)}, 

  title={Quantum Computing Education for Computer Science Students: Bridging the Gap with Layered Learning and Intuitive Analogies}, 

  year={2024},

  volume={03},

  number={},

  pages={61-70},

  doi={10.1109/QCE60285.2024.20460}
}

@INPROCEEDINGS{9951184,
  author={Temporão, Guilherme P. and Guerreiro, Thiago B. S. and Ripper, Pedro S. C. and Pavani, Ana M. B.},
  booktitle={2022 IEEE International Conference on Quantum Computing and Engineering (QCE)}, 
  title={Teaching Quantum Computing without prerequisites: a case study}, 
  year={2022},
  volume={},
  number={},
  pages={673-676},
  doi={10.1109/QCE53715.2022.00090}}

@ARTICLE{Carrascal2021-fl,
  title     = "First experiences of teaching quantum computing",
  author    = "Carrascal, Gin{\'e}s and del Barrio, Alberto A and Botella,
               Guillermo",
  journal   = "J. Supercomput.",
  publisher = "Springer Science and Business Media LLC",
  volume    =  77,
  number    =  3,
  pages     = "2770--2799",
  month     =  mar,
  year      =  2021,
  language  = "en"
}

@misc{LMU2025,
  title = "{Introduction to Quantum Computing}",
  howpublished = "https://www.nm.ifi.lmu.de/teaching/Vorlesungen/2025ss/quantum-computing/",
  note = "[Online; accessed 01-April-2025]",
  organization = "Ludwig Maximilian University of Munich"
}

@ARTICLE{Gangapuram2024-ot,
  title     = "Benchmarking quantum computer simulation software packages:
               State vector simulators",
  author    = "Gangapuram, Amit Jamadagni and L{\"a}uchli, Andreas and Hempel,
               Cornelius",
  journal   = "SciPost Phys. Core",
  publisher = "Stichting SciPost",
  volume    =  7,
  number    =  4,
  month     =  nov,
  year      =  2024,
  copyright = "https://creativecommons.org/licenses/by/4.0"
}

@misc{qiskit2024,
      title={Quantum computing with {Q}iskit},
      author={Javadi-Abhari, Ali and Treinish, Matthew and Krsulich, Kevin and Wood, Christopher J. and Lishman, Jake and Gacon, Julien and Martiel, Simon and Nation, Paul D. and Bishop, Lev S. and Cross, Andrew W. and Johnson, Blake R. and Gambetta, Jay M.},
      year={2024},
      doi={10.48550/arXiv.2405.08810},
      eprint={2405.08810},
      archivePrefix={arXiv},
      primaryClass={quant-ph}
}

@book{CirqDevelopers_2025, title={Cirq}, url={https://zenodo.org/doi/10.5281/zenodo.4062499}, DOI={10.5281/ZENODO.4062499}, abstractNote={Python package for writing, manipulating, and running quantum circuits on quantum computers and simulators.}, publisher={Zenodo}, author={Cirq Developers}, year={2025}, month=apr }

@article{openqasm3,
author = {Cross, Andrew and Javadi-Abhari, Ali and Alexander, Thomas and De Beaudrap, Niel and Bishop, Lev S. and Heidel, Steven and Ryan, Colm A. and Sivarajah, Prasahnt and Smolin, John and Gambetta, Jay M. and Johnson, Blake R.},
title = {OpenQASM 3: A Broader and Deeper Quantum Assembly Language},
year = {2022},
issue_date = {September 2022},
publisher = {Association for Computing Machinery},
address = {New York, NY, USA},
volume = {3},
number = {3},
url = {https://doi.org/10.1145/3505636},
doi = {10.1145/3505636},
journal = {ACM Transactions on Quantum Computing},
month = sep,
articleno = {12},
numpages = {50}
}

@Article{numpy,
 title         = {Array programming with {NumPy}},
 author        = {Charles R. Harris and K. Jarrod Millman and St{\'{e}}fan J.
                 van der Walt and Ralf Gommers and Pauli Virtanen and David
                 Cournapeau and Eric Wieser and Julian Taylor and Sebastian
                 Berg and Nathaniel J. Smith and Robert Kern and Matti Picus
                 and Stephan Hoyer and Marten H. van Kerkwijk and Matthew
                 Brett and Allan Haldane and Jaime Fern{\'{a}}ndez del
                 R{\'{i}}o and Mark Wiebe and Pearu Peterson and Pierre
                 G{\'{e}}rard-Marchant and Kevin Sheppard and Tyler Reddy and
                 Warren Weckesser and Hameer Abbasi and Christoph Gohlke and
                 Travis E. Oliphant},
 year          = {2020},
 month         = sep,
 journal       = {Nature},
 volume        = {585},
 number        = {7825},
 pages         = {357--362},
 doi           = {10.1038/s41586-020-2649-2},
 publisher     = {Springer Science and Business Media {LLC}},
 url           = {https://doi.org/10.1038/s41586-020-2649-2}
}

@misc{ibm_quantum_learning,
  author = {{IBM}},
  title = {{IBM Quantum Learning}},
  year = {2025},
  month = {04},
  howpublished = {\url{https://learning.quantum.ibm.com}},
  note = {\texttt{[Online]}}
}

@misc{zuse_tn_workshop,
  author = {{Zuse Institute Berlin}},
  title = {{Workshop on Tensor Methods for Quantum Simulation 2024}},
  year = {2024},
  month = {06},
  howpublished = {\url{https://tmqs2024.zib.de/}},
  note = {\texttt{[Online]}}
}

@misc{eu_tn_workshop_phd,
  author = {{European Tensor Network}},
  title = {{Tensor Network based approaches to Quantum Many-Body Systems}},
  year = {2024},
  month = {05},
  howpublished = {https://nextcloud.tfk.ph.tum.de/etn/index.php/schools/2024-school/},
  note = {\texttt{[Online]}}
}

@misc{la_hamburg,
  author = {{Uni Hamburg}},
  title = {{Lineare Algebra für Informatiker und Statistiker}},
  year = {2025},
  month = {04},
  howpublished = {https://www.math.uni-hamburg.de/home/belgun/LA-lmu.pdf},
  note = {\texttt{[Online]}}
}

@misc{la_tum,
  author = {{Kemper, Gregor}},
  title = {{Lineare Algebra für Informatik}},
  year = {2024},
  month = {10},
  howpublished = {\url{https://www.math.cit.tum.de/fileadmin/w00ccg/algebra/people/kemper/lectureNotes/LA_Inf.pdf}},
  note = {\texttt{[Online]}}
}

@misc{la_mit,
  author = {{Strang, Gilbert}},
  title = {{Linear Algebra}},
  year = {2023},
  month = {06},
  howpublished = {https://ocw.mit.edu/courses/18-06-linear-algebra-spring-2010/pages/syllabus/},
  note = {\texttt{[Online]}}
}

@misc{deeplearing_lecture,
  author = {{quantstart}},
  title = {{Scalars, Vectors, Matrices and Tensors - Linear Algebra for Deep Learning}},
  year = {2025},
  month = {04},
  howpublished = {https://www.quantstart.com/articles/scalars-vectors-matrices-and-tensors-linear-algebra-for-deep-learning-part-1/},
  note = {\texttt{[Online]}}
}

@ARTICLE{SycamoreExperiment,
  title     = "Quantum supremacy using a programmable superconducting processor",
  author    = "Arute, Frank and Arya, Kunal and Babbush, Ryan and Bacon, Dave
               and Bardin, Joseph C and Barends, Rami and Biswas, Rupak and
               Boixo, Sergio and Brandao, Fernando G S L and Buell, David A and
               Burkett, Brian and Chen, Yu and Chen, Zijun and Chiaro, Ben and
               Collins, Roberto and Courtney, William and Dunsworth, Andrew and
               Farhi, Edward and Foxen, Brooks and Fowler, Austin and Gidney,
               Craig and Giustina, Marissa and Graff, Rob and Guerin, Keith and
               Habegger, Steve and Harrigan, Matthew P and Hartmann, Michael J
               and Ho, Alan and Hoffmann, Markus and Huang, Trent and Humble,
               Travis S and Isakov, Sergei V and Jeffrey, Evan and Jiang, Zhang
               and Kafri, Dvir and Kechedzhi, Kostyantyn and Kelly, Julian and
               Klimov, Paul V and Knysh, Sergey and Korotkov, Alexander and
               Kostritsa, Fedor and Landhuis, David and Lindmark, Mike and
               Lucero, Erik and Lyakh, Dmitry and Mandr{\`a}, Salvatore and
               McClean, Jarrod R and McEwen, Matthew and Megrant, Anthony and
               Mi, Xiao and Michielsen, Kristel and Mohseni, Masoud and Mutus,
               Josh and Naaman, Ofer and Neeley, Matthew and Neill, Charles and
               Niu, Murphy Yuezhen and Ostby, Eric and Petukhov, Andre and
               Platt, John C and Quintana, Chris and Rieffel, Eleanor G and
               Roushan, Pedram and Rubin, Nicholas C and Sank, Daniel and
               Satzinger, Kevin J and Smelyanskiy, Vadim and Sung, Kevin J and
               Trevithick, Matthew D and Vainsencher, Amit and Villalonga,
               Benjamin and White, Theodore and Yao, Z Jamie and Yeh, Ping and
               Zalcman, Adam and Neven, Hartmut and Martinis, John M",
  journal   = "Nature",
  publisher = "Springer Science and Business Media LLC",
  volume    =  574,
  number    =  7779,
  pages     = "505--510",
  month     =  oct,
  year      =  2019,
  language  = "en"
}

@article{PhysRevLett.129.090502,
  title = {Solving the Sampling Problem of the Sycamore Quantum Circuits},
  author = {Pan, Feng and Chen, Keyang and Zhang, Pan},
  journal = {Phys. Rev. Lett.},
  volume = {129},
  issue = {9},
  pages = {090502},
  numpages = {6},
  year = {2022},
  month = {Aug},
  publisher = {American Physical Society},
  doi = {10.1103/PhysRevLett.129.090502},
  url = {https://link.aps.org/doi/10.1103/PhysRevLett.129.090502}
}

@ARTICLE{Boixo2018-ai,
  title     = "Characterizing quantum supremacy in near-term devices",
  author    = "Boixo, Sergio and Isakov, Sergei V and Smelyanskiy, Vadim N and
               Babbush, Ryan and Ding, Nan and Jiang, Zhang and Bremner,
               Michael J and Martinis, John M and Neven, Hartmut",
  journal   = "Nat. Phys.",
  publisher = "Springer Science and Business Media LLC",
  volume    =  14,
  number    =  6,
  pages     = "595--600",
  month     =  jun,
  year      =  2018,
  language  = "en"
}

@online{LRZQuantum,
  title        = {Quantum Integration Centre des LRZ },
  author       = {Leibniz-Rechenzentrum},
  year         = 2025,
  url          = {https://www.lrz.de/technologien/quantum/},
  urldate      = {04.15.2025}
}

\end{document}